\documentclass[12pt]{article}  
\usepackage{eufrak} 
\usepackage{latexsym}
\usepackage[mathscr]{eucal}

\newcommand{\g}{{\mathfrak g}}
\newcommand{\M}{{\mathfrak M}}
\newcommand{\X}{{\mathfrak X}}
\newcommand{\V}{{\mathfrak V}}

\newcommand{\E}{{\mathcal E}}
\newcommand{\A}{{\mathcal A}}
\newcommand{\B}{{\mathcal B}}

\newcommand{\Z}{{\mathbb Z}}

\newcommand{\Lie}{{\mathcal L}}
\def\beq{\begin{equation}}
\def\eeq{\end{equation}}
\def\beqa{\begin{eqnarray}}
\def\eeqa{\end{eqnarray}}
\font\mybb=msbm10 at 12pt
\font\mybbb=msbm10 at 8pt
\def\bbb#1{\hbox{\mybbb#1}}
\def\bb#1{\hbox{\mybb#1}}
\def\Z{\bb{Z}}
\def\C{\bb{C}}
\def\H{\bb{H}}
\def\bC{\bbb{C}}
\def\bR{{\bbb{R}}}

\def\real{{\bb{R}}}

\def\Ker{\mbox{Ker}}
\def\Tr{\mbox{\rm Tr}}
\def\uno{\mbox{1 \kern-.59em {\rm l}}}

\begin{document}
\begin{titlepage}
\begin{flushright}
{ROM2F/2000/25}\\
{SISSA 74/2000/fm}
\end{flushright}
\begin{center}
 
{\large \sc On the Multi-Instanton Measure for Super Yang--Mills Theories
}\\ 
  
\vspace{0.2cm}
{\sc Ugo Bruzzo}\\
{\sl Scuola Internazionale Superiore di Studi Avanzati,\\
Via Beirut 4, 34013 Trieste, Italy}\\
{\sc Francesco Fucito, Alessandro Tanzini}\\
{\sl Dipartimento di Fisica, Universit\'a di Roma ``Tor Vergata'',\\
I.N.F.N. Sezione di Roma II,\\
Via della Ricerca Scientifica, 00133 Roma, Italy}\\
and\\
{\sc Gabriele Travaglini}\\
{\sl Department of Physics, University of Durham,\\
Durham, DH1 3LE, UK}
\end{center}
\vskip 0.5cm
\begin{center}
{\large \bf Abstract}
\end{center}
{In this paper we revisit the
arguments that have led to the proposal of a
multi-instanton measure for supersymmetric Yang-Mills theories. 
We then recall how the moduli space of gauge connections on $\real^4$
can be built from a hyperk\"ahler quotient construction which we 
generalize to supermanifolds.
The measure we are looking for is given by the supermetric of the supermoduli
space thus introduced. To elucidate the construction
we carry out explicit computations in the case of $N=2$ 
supersymmetric Yang-Mills theories.}
    
\vfill
\end{titlepage}
\addtolength{\baselineskip}{0.3\baselineskip} 
\setcounter{section}{0}
\section{Introduction}                   
Great progresses have been made in recent years in the understanding
of multi-instanton calculus. A first impulse to try to perform
computations for winding numbers, $k$, bigger than one came from
the solution for the holomorphic part of the effective 
action for extended globally $N=2$ supersymmetric Yang-Mills (SYM) 
theories  proposed in ref.\cite{sw}. The idea of checking this solution 
triggered
a first set of computations up to $k=2$ \cite{fp, DKM, ft} in the
case with no matter. Computations for $k>2$ seemed to be
out of reach due to the lack of an explicit parametrization for
the ADHM data \cite{adhm} and for the complexity of the algebra 
involved in the computation. Trying to circumvent this limitation,
in ref.\cite{DKM1} a new computational strategy was devised: the 
ADHM constraints were inserted in the path integral through the
introduction of a certain number of Dirac deltas. This removed the
need of solving complicated algebraic equations of higher order.
The product of the Dirac deltas times the differentials of the
fermionic and bosonic zero--modes involved in the computation
is the supermeasure that needs to be specified to perform the
computation. 
The Dirac deltas needed to implement the constraints were found
working backwards i.e. starting from the explicit
form of the measure, known in the $k=2$ case \cite{osb}.
This approach was then extended to arbitrary $k$. 
The measure thus obtained possesses a certain number of desired
features which are dictated by physical considerations:
it is supersymmetric, reproduces the known $k=2$ measure 
for $N=1,2$ and in
the dilute gas limit factorizes as expected. The extension of this
procedure to the $N=4$ case was also given \cite{DKM2}.

Although this procedure may seem to be rather {\it ad hoc}, 
it has anyway proved to be very useful in giving a
nonperturbative
consistency check of the conjectured duality \cite{mal} between
certain IIB string theory correlators on an  
$AdS_5\times S^5$ background and some
Green's functions of composite operators of the
$N=4$  $SU(N_c )$ SYM theory in four dimensions in the large 
$N_c$ limit. At leading order in 
that limit the $N=4$ measure, for moduli spaces of
dimension $k$, collapses to the product of
the measure of an $AdS_5\times S^5$ space times the partition
function
of an $N=1$ ten dimensional $SU(k)$ theory reduced to
0+0 dimensions \cite{DKM3}.
It was then possible to carry out an
explicit calculation which turned out to be in agreement with the
above mentioned conjecture \cite{DKM3}: a highly nontrivial 
result, in our opinion.

Another motivation for the present study comes from 
ref.\cite{bftt}, where multi--instanton calculus was 
reformulated in the language of topological field theories.
This gives a new geometrical interpretation of the
nonperturbative effects and from a computational point of view
allows to rewrite all the correlators of interest as total
derivatives on the  moduli space of gauge connections. This, in turn,
could leads to further progresses in the computations for generic 
values of $k$'s,
using the properties of the ADHM contruction at the boundary 
of the moduli space. A key ingredient in this approach is given by
the introduction of a derivative on the moduli space of gauge connections:
the nilpotent BRST operator of the theory, $s$. 
If the moduli space is
realized via the ADHM construction, the nilpotency of $s$ requires
the introduction of a connection which can be explicitly
computed imposing the fermionic constraint of the ADHM 
construction \cite{bftt} or, alternatively, from
the Killing vectors of the residual symmetry of the above mentioned
constraints. 
Furthermore, the BRST equations relate bosonic differentials with
fermionic variables through the above cited connection, 
which consequently appears
in the multi--instanton measure, as computed in ref.\cite{bftt},
by this change of basis.
In view of the potential applications of these 
results, the scope of this paper is to revisit the derivation of the
supersymmetric instanton measure in the light of the geometry of
the moduli space of gauge connections. By generalizing the standard
bosonic construction for the metric of a hyperk\"ahler quotient
to its supersymmetric extension, we will derive the measure from
this newly introducted supermetric. The construction is valid for
any $k$, though it can be made explicit only for $k=2$. For higher
$k$'s we will introduce Dirac deltas, following ref.\cite{DKM1},
which will directly implement the necessary steps of the quotient
construction.
This will allow us to put the derivation of the measure on 
a firmer mathematical basis than before
since the implemented constraints do not have their origin only in
symmetry arguments but stem from the quotient construction.

The plan of this paper is the following: in the next section 
after introducing the quotient construction, we give a brief presentation
of the ADHM construction in this language and recall how to compute
the bosonic metric of the moduli space in the $k=2$ case. In the
third section we recall some material from ref.\cite{bftt} we need here.
In the fourth section we give the supersymmetric extension of the
quotient construction.
Finally in the fifth and last section we show how to implement the
construction of section \ref{quattro} in the functional integral
giving the nonperturbative contribution for arbitrary 
winding number $k$. 

\section{The Hyperk\"ahler Quotient Construction\label{quotients}}
\setcounter{equation}{0} 
We start
by discussing the Marsden-Weinstein reduction \cite{ms} which
allows us to define a metric
$\tilde g$ on the quotient $M=V/G$, where
$G$ is a Lie group with Lie algebra $\g$ which acts by isometries
on a Riemannian manifold $(V,g)$. An alternative discussion of this
construction, using nonlinear sigma models, can be found in
ref.\cite{hklr}. An application to connections on gravitational instantons
is given in ref.\cite{bfmr}.
Let us assume that the action of $G$ is proper and
free. Then the quotient $M=V/G$ is a smooth manifold, and the projection
$\pi\colon V\to M$ is a principal bundle with structure group $G$.
For every $\xi\in \g$ the associated infinitesimal generator $\xi^\ast$ of the
action of $G$ on $V$ is a vertical fundamental vector field for the
principal bundle $V$. 

For every $x\in V$ the collection $\{\xi^\ast(x)\}_{\xi\in\g}$ coincides with
the vertical tangent space $\mbox{Vert}_xV$. If we set
$\mbox{Hor}_xV=(\mbox{Vert}_xV)^\perp$ in the metric
$g$, the assignment $x\mapsto \mbox{Hor}_xV$ is $G$-equivariant and therefore
defines a connection on $V$. We shall denote by $C$ the corresponding
connection form (as a $\g$-valued form on $V$). We have now
an associated
\emph{horizontal lift operator:} for every vector field $\alpha$ on $M$, its
horizontal lift $\tilde\alpha$ is the unique $G$-invariant vector field on $V$
which projects to $\alpha$.

The metric $g$ induces a metric $\tilde g$ on $M$, given by
\begin{equation}
\tilde g(\alpha,\beta)=g(\tilde\alpha,\tilde\beta).
\label{uno}
\end{equation}
Let us write $\tilde g$ in components. Given local coordinates
$(y^1,\dots,y^m)$ in $M$ and a basis $\{\xi_a\}$ of $\g$ we may represent
the horizontal lift in the form
\begin{equation}
\widetilde{\frac{\partial}{\partial y^i}}=\frac{\partial}{\partial
y^i}-C_i^a\,\xi^\ast_a.
\label{due}
\end{equation}
Given the definition (\ref{due}) for the horizontal lift
and keeping in account that $\{\xi^\ast(x)\}_{\xi\in\g}$
coincides with the vertical tangent space,
the following identity holds
\begin{equation}
\label{identity}
0 = g\left(\widetilde{\frac{\partial}{\partial y^i}},\xi_a^\ast\right)=
g\left(\frac{\partial}{\partial
y^i},\xi_a^\ast\right)-C_i^b\,g(\xi_b^\ast,\xi_a^\ast).
\end{equation}
The matrix $g_{ab}=g(\xi_b^\ast,\xi_a^\ast)$ is invertible;
by denoting by $g^{ab}$ the elements of the inverse matrix, we get
\begin{equation}
C_i^a=g^{ab}\,g\left({\frac{\partial}{\partial
y^i}},\xi_b^\ast\right).
\label{tre}
\end{equation}
Acting with the metric $g$ on two elements of the 
horizontal lift (\ref{due}), we get
\begin{equation}
\label{metric}
\tilde g_{ij}= g_{ij}-g^{ab}
g\left({\frac{\partial}{\partial
y^i}},\xi_a^\ast\right) 
g\left({\frac{\partial}{\partial
y^j}},\xi_b^\ast\right)=g_{ij}-C_i^aC_{aj}.
\end{equation}

It is now possible to define a hyperk\"ahler quotient in the
following way:
let $X$ be a hyperk\"ahler manifold of real dimension $4n$, with hyperk\"ahler
metric $g$ and basic complex structures $J_i$, $i=1,2,3$. Let $\omega_i$ be
the corresponding K\"ahler forms. Assume
that a Lie group $G$ acts on $X$ freely and properly
by hyperk\"ahler isometries, so that
\begin{equation}
\Lie_\xi\omega_i=0,
\label{quatro}
\end{equation}
for all $\xi\in\g$ (here $\Lie$ is the Lie derivative).
As a result, provided that $H^1(X,\real)=0$, 
and having fixed a basis $\{\xi_a\}$ of $\g$, one gets $3r$ ``first integrals''
$f^a_i$
($r=\dim G$) such that 
\begin{equation}
0= \Lie_{\xi_a}\omega_i=df_i^a.
\label{cinque}
\end{equation}
Let $V$ be the submanifold of $X$ defined by the equations
$f^a_i=0$, so that $\dim V=4n-3r$. The group $G$ acts freely
and properly on $V$, and one has a quotient $M=V/G$
of dimension $4(n-r)$. Every complex structure
on $X$, compatible with $g$, defines a complex structure on $M$, 
and one can prove that the quotient metric $\tilde g$ is hyperk\"ahler. We can now bridge this construction with the
standard ADHM one.

The starting point is the ADHM matrix $\Delta=a+bx$. Due to
the symmetries of the ADHM contruction (we will later come back to
this point at greater length) we may choose the matrix $b$  
so that it does not contain any moduli. Then in the case of the gauge group
$SU(n)$, the matrix $a$ can be written as
\beq
\label{salute1}
a=\pmatrix{t & s^\dagger \cr A & -B^\dagger\cr B & A^\dagger} \ \ ,
\eeq
where 
$A, B$ are  $k\times k$ complex matrices and $s, t$ are $n\times k$
and $k\times n$ dimensional matrices.  
Let us introduce the  $4k^2+4kn$--dimensional 
hyperk\"{a}hler manifold $M=\{A,B,s,t\}$. 
Given the three complex structures
$J^i_{ab}$ where $i=1,2,3$ and $a, b=1,\ldots,{\rm dim}\, M$, 
we can build the 2--forms
$\omega^i=J^i_{ab}dx^a\wedge dx^b$,  where the $x^a$'s are coordinates
on $M$. The real forms $\omega^i$ allow one to define a $(2,0)$ 
and a $(1,1)$ form
\beqa
\label{comforms}
\omega_{\bC}&=& \Tr\, dA\wedge dB+\Tr\, ds\wedge dt\ \ ,\nonumber\\
\omega_{\bR}&=&\Tr\, dA\wedge dA^\dagger+\Tr\, dB\wedge dB^\dagger+
\Tr\, ds\wedge ds^\dagger- \Tr\, dt^\dagger\wedge dt\ \ .
\eeqa
The transformations
\beqa\label{invmod}
A&\rightarrow& QA Q^\dagger\ \ ,\nonumber\\
B&\rightarrow& QB Q^\dagger\ \ ,\nonumber\\
s&\rightarrow& Qs R^\dagger\ \ ,\nonumber\\
t&\rightarrow& Rt Q^\dagger\ \ ,
\eeqa
with $Q\in U(k), R\in U(n)$ leave $\omega_{\bC}, \omega_{\bR}$ invariant.
Using a complex notation for the
momenta defined in (\ref{cinque}) $f^i_\xi=f^i_a\xi^a$,
we write 
\beqa
f_{\bC}&=&[A,B]+st\ \ ,
\nonumber\\
f_{\bR}&=&[A,A^\dagger]+[B,B^\dagger]+ss^\dagger-t^\dagger t
\ \ .
\eeqa
$f^i_\xi=0$ defines a hypersurface $\mathscr{N}^{+}$ in $M$,
of dimension $k^2+4kn$.
The moduli space of self--dual gauge
connections, $\mathscr{M}^{+}$, is obtained by 
modding $\mathscr{N}^{+}$ by the reparametrizations  
defined in (\ref{invmod}). It has dimension 
${\rm dim}\,\mathscr{M}^{+}=4kn$ and, as we have already noticed, is 
hyperk\"ahler. To make things even more explicit and as a guidance
for future developments we now explicitly perform the $k=2$
computation in the $SU(2)$ case \cite{bftt}. In order to do this,
we pause to adapt our notation to this case.
In fact we find it convenient to introduce a quaternionic notation
exploiting the isomorphism between  $SU(2)$ and $Sp(1)$.
The points, 
$x$, of the 
quaternionic space $\H\equiv \C^2\equiv\real^4$ 
can be conveniently
represented in the form $x=x^\mu \sigma_\mu$, with 
$\sigma_\mu=(i\sigma_c , \uno_{2\times 2}), c=1,2,3.$ 
The $\sigma_c$'s 
are the usual Pauli matrices, and
$\uno_{2\times 2}$ is the 2--dimensional identity matrix.
The conjugate of $x$ is $x^{\dagger} = 
x^\mu \bar{\sigma}_\mu$. 
A quaternion is said to be real if it is proportional to 
$\uno_{2\times 2}$ 
and imaginary if it has vanishing real part. 

The prescription to find an instanton of winding number $k$ is 
the following: introduce a $(k+1)\times k$ quaternionic matrix
linear in $x$
\beq
\Delta=a+bx \  ,
\label{f.4}
\eeq
where $a$ has the generic form 
\beq
\label{salute}
a=\pmatrix{w_1&\ldots&w_k\cr{}&{}&{}\cr{}&a'&{}\cr{}&{}&{}}\ \ ;
\eeq
$a'$ is a $k\times k$ quaternionic matrix.
The (anti--hermitean) gauge connection  takes the form 
\beq
A=U^\dagger d U
\ \ ,
\label{trecinque}
\eeq
where $U$ is a $(k+1)\times 1$ matrix of quaternions providing an
orthonormal frame of $\Ker \Delta^\dagger$, {\it i.e.}
\beqa
\Delta^\dagger U &=& 0
\ \ ,
\label{f.5}
\\
U^\dagger U &=&\uno_{2\times 2} 
\ \ .
\label{f.6}
\eeqa
The constraint (\ref{f.6}) ensures that $A$ 
is an element  of the Lie algebra of the $SU(2)$ gauge group.
The self--duality condition 
\beq
\label{ddd}
{}^{\ast} F = F
\eeq
on the field strength of the gauge connection (\ref{trecinque}) 
requires the matrix $\Delta$ 
to obey the constraint
\beq
\label{bos}
\Delta^\dagger\Delta=(\Delta^{\dagger}\Delta)^{T}
\ \ ,
\eeq
where the superscript $T$ stands for transposition of the
quaternionic
elements of the matrix (without transposing the quaternions 
themselves).
(\ref{bos}) in turn implies 
$\Delta^\dagger\Delta=f^{-1}\otimes\uno_{2\times 2}$, 
where $f$ is an invertible hermitean $k\times k$ matrix 
(of real numbers). 

Gauge transformations are implemented in this formalism 
as  right multiplication   of $U$
by a unitary (possibly $x$--dependent) quaternion.
Moreover, $A$ is invariant  under 
reparametrizations of the ADHM data of the form:
\beq
\label{nonbanane}
\Delta\rightarrow  Q\Delta R \ ,
\eeq
with $Q\in Sp(k+1), R\in GL(k,\real)$.  It is straightforward to see 
that (\ref{nonbanane}) preserves the bosonic constraint (\ref{bos}).
These symmetries  can be used to simplify the expressions
of $a$ and $b$. Exploiting this fact, in the
following we will 
choose the matrix $b$ to be
\beq
b=-\pmatrix {0_{1\times k}\cr\uno_{k\times k}}.
\label{boh}
\eeq
Choosing the canonical form (\ref{boh}) for $b$,
the bosonic constraint (\ref{bos}) becomes
\beqa
\label{a.1}
&&a^{\prime} = {a^\prime}^{T} \ \ ,
\\
&&a^\dagger a = (a^\dagger a  )^T \ \ .
\label{a.2}
\eeqa
This 
still allows for $O(k)\times SU(2)$ reparametrizations 
of the form
(\ref{nonbanane}),  where now $R\in O(k)$,  
\beq\label{ahoo}
Q = \pmatrix{q&0&\ldots&0\cr
0&{}&{}&{}\cr \vdots&{}&R^T&{}\cr
0&{}&{}&{}}
\ \ , 
\eeq
and 
$q\in SU(2)$.
These transformations act nontrivially on 
the matrix $a$ and leave $b$  invariant. 
After imposing the constraint (\ref{bos}), the number of independent 
degrees of freedom contained in $\Delta$  
(that is the number of independent collective coordinates 
that the ADHM formalism uses to describe an instanton of winding number $k$)
is $8k + k(k-1) / 2$; modding  out  the $O(k)\times SU(2)$ 
reparametrization transformations, we remain with $8k-3$ 
independent degrees of freedom. However (\ref{f.5}) and (\ref{f.6})
do not determine $U_0/ |U_0|$, where $U_0$ is the first component 
of $U$; this adds three extra degrees of freedom, so that in conclusion 
we end up with  a moduli space of dimension $8k$
(the instanton moduli space $\mathscr{M}^{+}$).
It is easy to convince oneself that the arbitrariness in $U_0/ |U_0|$
can be traded for the $SU(2)$ reparametrizations; in other words,
one can forget to mod out the $SU(2)$ factor of 
the reparametrization group $O(k)\times SU(2)$
but fix the phase of the quaternion $U_0$ (setting  for example
$U_0 = |U_0|\uno_{2\times 2}$). 
This is what we will actually do in the following.

We now focus our attention on the zero--modes corresponding to
the self--dual field introduced in (\ref{trecinque}).
They must obey \cite{osb}
\beq
{}^{\ast}( D_{[\mu} \psi_{\nu ]} ) = D_{[\mu} \psi_{\nu ]},\quad 
 D_{\mu} \psi_{\mu}=0 \ \ , 
\label{f.8}
\eeq
where $D$ is the covariant derivative in the instanton 
background,  Eq.(\ref{trecinque}).
The solution to (\ref{f.8}) can be written as  \cite{osb}
\beq
\label{f.9}
\psi=U^{\dagger}{\cal M}f(d\Delta^{\dagger})U+
U^{\dagger}(d\Delta)f{\cal M}^{\dagger}U\ \ , 
\eeq
where ${\cal M}$ is a $(k+1)\times k$ matrix of quaternions which,
in order 
for (\ref{f.8}) to be satisfied  must obey the
constraint
\beq
\label{fconstr}
\Delta^{\dagger}{\cal M}=(\Delta^{\dagger}{\cal M})^T \ \ .
\eeq
(\ref{f.8}) tell us that the $\psi$ zero--modes are the tangent vectors 
to the instanton moduli space $\mathscr{M}^+$; as it is well known, 
the number of independent zero--modes  is $8k$ (the
dimension of $\mathscr{M}^+$), and we would like to see how 
this is realized
in the formalism of the ADHM construction. To this end, 
note that 
${\cal M}$ has 
$k(k+1)$ quaternionic elements ($4k(k+1)$ real degrees of freedom)
which are subject to the
$4k(k-1)$ constraints given by (\ref{fconstr}). The number of independent 
${\cal M}$'s 
satisfying (\ref{fconstr}) is thus $8k$,  as desired.
Notice that since the $O(k)$ symmetry is purely bosonic it
only mods out the degrees of freedom in $a$, leaving those
in ${\cal M}$ untouched. This fact will play an important role
in the following.

If we work in  the gauge where $b$ has the canonical form
(\ref{boh}), then (\ref{fconstr}) can be conveniently elaborated as follows.
We put ${\cal M}$ in a form which parallels that for $a$ in (\ref{salute}),
\beq
\label{salutem}
{\cal M}=\pmatrix{\mu_1&\ldots&\mu_k\cr{}&{}&{}\cr{}&{\cal M}'&{}\cr{}&{}&{}}
\ \ , 
\eeq
${\cal M}^\prime$ being a $k\times k$ quaternionic matrix.
Plugging  (\ref{salute}), (\ref{boh}), (\ref{salutem})
into (\ref{fconstr}) we get
\beqa
\label{m.1}
&&{\cal M}^{\prime} = {{\cal M}^\prime}^{T} \ \ ,
\\
&&a^\dagger {\cal M} = (a^\dagger {\cal M}  )^T \ \ .
\label{m.2}
\eeqa
When $\Delta$ is transformed according to (\ref{nonbanane}),
the ${\cal M}$'s must also be reparametrized so as to keep 
the constraint (\ref{fconstr})  unchanged. This implies that 
the ${\cal M}$'s must undergo the same formal reparametrization as
$\Delta$, that is 
\beq
\label{nonbananen}
{\cal M}
\rightarrow  Q{\cal M} R
\ \ .
\eeq

Let us now consider the $k=2$ case explicitly.
The ADHM bosonic matrix reads
\beq
\Delta=\pmatrix{w_1 & w_2\cr x_1 - x & a_1\cr a_1 & x_2 - x}
=\pmatrix{w_1 & w_2\cr a_3 & a_1\cr a_1 & -a_3}+b(x-x_0) \ \ ,
\label{f.14}
\eeq
where $x_0=(x_1+x_2)/2$, $a_3=(x_1-x_2)/2$. 
We also need the expression of the matrix ${\cal M}$ 
which is defined in (\ref{fconstr}). Since this constraint 
is very similar to (\ref{bos}) (to get convinced of 
this fact just think that two solutions of (\ref{fconstr}) are given 
by ${\cal M}$ proportional to $a$ and $b$, respectively)
it is convenient to choose a form of ${\cal M}$ which parallels 
(\ref{f.14})
\beq
{\cal M}=\pmatrix{\mu_1 & \mu_2\cr \xi+{\cal M}_3 & {\cal M}_1\cr {\cal M}_1
& \xi-{\cal M}_3}
=\pmatrix{\mu_1 & \mu_2\cr {\cal M}_3& {\cal M}_1\cr 
{\cal M}_1& -{\cal M}_3}- b\xi\ \ .
\label{f.155}
\eeq
The bosonic constraint (\ref{bos}) now reads  
\beq 
\label{natale1}
\bar{w}_2 w_1-\bar{w}_1w_2=2(\bar{a}_3 a_1-\bar{a}_1a_3),
\eeq
(\ref{natale1}) is a set of three equations since both sides 
are purely imaginary.
We decide to solve (\ref{natale1}) with respect to $a_1$.
A possible solution is 
\beq 
\label{natale}
a_1=\frac{1}{4|a_3|^2}a_3(\bar{w}_2 w_1-\bar{w}_1w_2 +\Sigma)\ \ ,
\eeq
where the imaginary part is fixed by (\ref{natale1}) and the free real
part has been called $\Sigma$.
It is easy to see that
\beq
\Sigma=2(\bar a_3 a_1+\bar a_1 a_3).
\label{snatale}
\eeq
The constraint (\ref{fconstr}) is 
\beq
\bar{w}_2 \mu_1-\bar{w}_1\mu_2=2(\bar{a}_3 M_1-\bar{a}_1M_3)\ \ ,
\eeq
and it is satisfied by
\beq
{\cal M}_1=\frac{a_3}{2|a_3|^2}(2\bar{a}_1 {\cal M}_3+
\bar w_2\mu_1-\bar w_1\mu_2)\ \ .
\label{f.16}
\eeq
As one can easily check, these are four real equations. The
dimension of the tangent space to the moduli space is the right
one without resorting to a quotient procedure.
Let us now introduce a 20--dimensional hyperk\"{a}hler manifold 
$M=\{w_1, w_2, a_3, a_1, x_0\}$.%
\footnote{Notice that, since we are
using a different parametrization of the ADHM space with respect to 
(\ref{salute1}), the dimensions of the manifolds $M$ and $\mathscr{N}^{+}$ 
are not those of the previous discussion.
However, also the reparametrization groups are different, in such a way 
that the final dimension of the moduli space of self--dual
gauge connections is the same, as it should be.}
The parametrization (\ref{f.14}) involves
the combination $x_0-x$ since the ADHM construction has a rigid
translation symmetry. We then find handy to restrict the analysis  
to the  16--dimensional hyperk\"{a}hler manifold
$M \backslash \{x_0\}$
parametrized by the quaternionic coordinates
\beq
m^I=(w_1,w_2,a_3,a_1),
\label{coordnv}
\eeq 
and endowed with a flat metric 
\beq
ds^2 = 
\eta_{I\bar J}dm^I d\bar m^{\bar J} = 
|dw_1|^2 + |dw_2|^2 + |da_3|^2 + |da_1|^2   
\ \ ,
\label{piatta}
\eeq
which, following \cite{hklr}, can be also imagined to be the
Lagrangian density of a suitable sigma model with target space 
$M \backslash \{x_0\}$.
To keep the notation 
as simple as possible,  we rename $\mathscr{M}^+\backslash \{x_0\}$ and
$\mathscr{N}^+\backslash \{x_0\}$ as 
$\mathscr{M}^+$, $\mathscr{N}^+$,
 respectively. 

(\ref{natale1}) is invariant under 
the reparametrization group $O(2)$, whose action on the $k=2$
quaternionic coordinates is
\beqa
&&(w_1^\theta , w_2^\theta) = (w_1 , w_2) R_\theta \ \ , \nonumber \\
&&(a_3^\theta , a_1^\theta) = (a_3 , a_1) R_{2\theta} \ \ ,
\label{ohdue}
\eeqa
with
\beq
R_\theta =
\pmatrix{\cos\theta & \sin\theta \cr -\sin\theta & \cos\theta} \ \ .
\label{rmatrix}
\eeq 
The construction of the reduced bosonic moduli space $\mathscr{M}^+$ 
proceeds now in two steps.
First, given the $O(2)$ invariant
solution (\ref{natale}) to the constraint (\ref{natale1}), 
$\mathscr{N}^{+}$ 
turns out to be a 13--dimensional manifold, 
described by the set of coordinates $(w_1,w_2,a_3,\Sigma)$. 
Second, we mod out the isometry group
of $\mathscr{N}^{+}$ as discussed above.
The instanton moduli space is then $\mathscr{M}^{+}=\mathscr{N}^{+}/O(2)$, 
and it has dimension 
${\rm dim}\,\mathscr{M}^{+}={\rm dim}\,\mathscr{N}^{+}- k(k-1)/2|_{k=2}=12$.
As anticipated, the construction of the quotient space 
$\mathscr{M}^{+}$ leads to the connection (\ref{tre})
which can also be obtained
by gauging a nonlinear sigma model \cite{hklr}. In this case we get
\beq
 C = {1\over{|k|^2}}\eta_{I\bar J}
    \bigl(\bar k^{\bar J} dm^I + d\bar m^{\bar J}k^I \bigr) \ \ ,
\label{pizza}
\eeq
where $k^I\partial_I + \bar k^{\bar I}\bar\partial_{\bar I}$ 
is the $O(k)$ Killing vector with 
$|k|^2 = \eta_{I\bar J} k^I \bar k^{\bar J}$.
The components of the $O(2)$ Killing vector on $M$
leaving  (\ref{natale}) invariant are 
\beq
k^I = (-w_2, w_1, -2a_1, 2a_3) \ \ .
\label{killing}
\eeq
Substituting (\ref{killing}) into (\ref{pizza}), we get 
\beqa
C &=& 
{1\over 2H}\Bigl(\bar w_1 dw_2 - \bar w_2 dw_1 + 
2\bar a_3da_1 - 2\bar  a_1da_3 + 
\nonumber\\ 
&& + d\bar w_2 w_1 -d\bar w_1 w_2 + 2d\bar a_1 a_3 -2d\bar a_3 a_1 \Bigr) 
\ \ .
\label{a}
\eeqa

The metric 
$g^{\mathscr{N}^{+}}_{I\bar J}$
on the constrained hypersurface $\mathscr{N}^{+}$
is obtained plugging (\ref{natale}) into (\ref{piatta}),  
and gets simplified if we introduce the variable 
\beq
W = \bar{w}_2 w_1 \ \ .
\label{wgrande}
\eeq
The hypersurface $\mathscr{N}^+$ is now described by the new set of 
coordinates
$(w_1, U, V, a_3, \Sigma)$, where 
\beqa
&& U = {{W + \overline W}\over 2} \ \ , \nonumber\\
&& V = {{W - \overline W}\over 2} 
\ \ ,
\label{uvu}
\eeqa
are respectively the real and the imaginary part of $W$. 
The Jacobian factor associated to this change of variables 
is 
\beq
d^4w_1 dU d^3V = |w_1|^4 d^4w_1 d^4w_2 
\ \ .
\label{iacobo}
\eeq
In the new variables, (\ref{piatta}) reads
\beqa 
ds^2 
&=& \Bigl(1+{{|w_2|^2}\over{|w_1|^2}}\Bigr)|dw_1|^2 + {dU^2\over{|w_1|^2}} 
   + {|dV|^2\over{|w_1|^2}} + \nonumber\\
&& - {dU\over{|w_1|^2}} (\bar w_2 dw_1 + d\bar w_1 w_2) 
   + {dV\over{|w_1|^2}} (\bar w_2 dw_1 - d\bar w_1 w_2) + \nonumber\\
&& + |da_3|^2 + |da_1|^2 \ \ ,
\label{flat}
\eeqa
which, inserting (\ref{natale}), becomes 
\beqa 
ds^2 &=& 
\Bigl(1+{{|w_2|^2}\over{|w_1|^2}}\Bigr)|dw_1|^2 + {dU^2\over{|w_1|^2}} 
   + {|dV|^2\over{|w_1|^2}} + \nonumber\\
&& - {dU\over{|w_1|^2}} (\bar w_2 dw_1 + d\bar w_1 w_2) 
   + {dV\over{|w_1|^2}} (\bar w_2 dw_1 - d\bar w_1 w_2) + \nonumber\\
&& + \Bigl(1+{{|a_1|^2}\over{|a_3|^2}}\Bigr)|da_3|^2      
   + {{d\Sigma}^2\over{16|a_3|^2}}
   + {|dV|^2\over{4|a_3|^2}} + \nonumber\\
&& - {{d\Sigma}\over{4|a_3|^2}} (\bar a_1 da_3 + d\bar a_3 a_1)
   - {dV\over{2|a_3|^2}} (\bar a_1 da_3 - d\bar a_3 a_1) 
\ \ .
\label{come} 
\eeqa
Also in this case,
the r.h.s. of (\ref{come}) can be regarded as the Lagrangian density of a 
zero--dimensional non--linear sigma model with target space 
$\mathscr{N}^{+}$.
In real coordinates 
\beq
m^A=(w_1^\mu, U, V^i, a_3^\mu, \Sigma),
\label{coordv}
\eeq 
the $O(2)$ Killing vector on this manifold has components
\beq
k^A = \Bigl(-w_2^\mu, |w_1|^2 - |w_2|^2, 0, -2a_1^\mu, 
8(|a_3|^2 - |a_1|^2)\Bigr) \ \ .
\label{kill}
\eeq
The global $O(2)$ symmetry can be promoted to a 
local one by introducing the connection (\ref{pizza}), 
which on $\mathscr{N}^{+}$ is written as 
\beqa
C &=&  {{g_{AB}^{\mathscr{N}^+}k^B}\over H}dm^A = \nonumber\\
&=& {1\over H}\Bigl(- 2w_2^\mu dw_1^\mu + dU - 4a_1^\mu da_3^\mu +
{{d\Sigma}\over 2}\Bigr) \ \ ,
\label{pizza1}
\eeqa
where the metric  
$g_{AB}^{\mathscr{N}^{+}}$ is obtained by rewriting (\ref{come})
in the coordinates $\{m_A\}$.
Writing $U$ in terms of $w_1, w_2$ by means of (\ref{wgrande}) 
and (\ref{uvu}),
the connection (\ref{pizza1}) becomes 
\beq
C = 
{1\over H}\Bigl( w_1^\mu dw_2^\mu -w_2^\mu dw_1^\mu - 4a_1^\mu da_3^\mu + 
{{d\Sigma}\over 2}\Bigr) \ \ .
\label{aripizza}
\eeq 
From (\ref{metric}), or alternatively from the 
gauged version of the Lagrangian given from (\ref{come}), 
we can read off the metric on $\mathscr{M}^{+} = 
\mathscr{N}^{+}/O(2)$ written in the $\{m^A\}$ coordinates, 
namely \cite{hklr}
\beq
g^{\mathscr{M}^{+}}_{AB}= g^{\mathscr{N}^{+}}_{AB} - 
{{g_{AC}^{\mathscr{N}^{+}}g_{BD}^{\mathscr{N}^{+}}k^Ck^D}
\over {g_{EF}^{\mathscr{N}^{+}}k^Ek^F}} 
\ \ .
\label{metric1}
\eeq 
We can now gauge fix the connection
by imposing $\Sigma=0$. This is a good gauge condition provided that
the hypersurface $\Sigma=0$ is transversal to the Killing vector
{\it i.e.} $d\Sigma(k)=|a_3|^2 - |a_1|^2\neq 0$ (using (\ref{kill})).

Finally, by using translational invariance to restore the 
dependence on $x_0$, and
taking into account the Jacobian factor (\ref{iacobo}), we 
write the volume form on the moduli space of self--dual gauge connections 
with winding number $k=2$ as  
\beq
|w_1|^4\sqrt{g^{\mathscr{M}^{+}}_{\Sigma=0}}
d^4w_1d^4w_2d^4a_3d^4x_0 =
{H\over{|a_3|^4}}\Big| |a_3|^2 - |a_1|^2 \Big| d^4w_1d^4w_2d^4a_3d^4x_0 
\ \ ,
\label{fine}
\eeq
which reproduces Osborn's well--known result \cite{osb}.

As we shall see in section \ref{quattro}, in terms of the cotangent
bundle to $\mathscr{M}^{+}$ we may construct a supermanifold whose
bosonic part is given by $\mathscr{M}^{+}$ itself.
The norm of the field $\psi$ will define the odd part of the supermetric
of the supermanifold.

The norm of the zero modes (\ref{f.9}) was computed 
by Corrigan to yield
\beqa
<\psi|\psi>&=&-\int_{\bR^4}|\psi|^2=2\pi^2\Tr \left[{\cal M}^\dagger
(1+P_\infty){\cal M} \right]=\eta_{I\bar J}M^I \bar M^{\bar J}\nonumber\\
& =& 
|\mu_1|^2 + |\mu_2|^2 + |M_3|^2 + |M_1|^2,
\label{kalformfer}
\eeqa
neglecting the
coordinate $\xi$ which is the "partner" of $x_0$.
We then introduce "rotated" variables \cite{DKM}
\beq
\mu_1={w_2\bar a_1 \mu_1^\prime\over |w_2||a_1|},\quad\quad
\mu_2={w_2\bar a_1 \mu_2^\prime\over |w_2||a_1|}.
\eeq 
Substituting (\ref{f.16}) into (\ref{kalformfer}) we get
\beqa
<\psi|\psi>&=& 
(1+{|w_2|^2\over 4|a^3|^2})(\mu_1^\prime)_\alpha(\mu_1^\prime)_\alpha + 
(1+{|w_1|^2\over 4|a^3|^2})(\mu_2^\prime)_\alpha(\mu_2^\prime)_\alpha + 
(1+{|a_1|^2\over |a_3|^2})(M_3)_\alpha(M_3)_\alpha +\nonumber \\
&&{|w_2||a_1|\over 2|a_3|^2}(\mu_1^\prime)_\alpha (M_3)_\alpha-   
{|w_1||a_1\over 2|a_3|^2}|(\mu_2^\prime)_\alpha (M_3)_\alpha-
{|w_1||w_2|\over 4|a_3|^2}(\mu_1^\prime)_\alpha 
(\mu_2^\prime)_\alpha ,
\label{piattatg1}
\eeqa
where $(M^I)_\alpha$ are the real components of the quaternion $M^I$.
From (\ref{kalformfer}), after restoring the $\xi$ dependence,
we easily compute
\beq
\sqrt{<\psi|\psi>}={H^2\over |a_3|^4}.
\eeq

\section{Connection with Topological Field Theories\label{topological}}
\setcounter{equation}{0}
In ref.\cite{bftt} it was shown how the results of instanton 
calculus can be more easily derived in the framework of topological
field theories \cite{witten}. Here we collect some results 
which will be relevant to our discussion.
We use the same notation as in \cite{bftt} to which we refer the reader
for a detailed exposition of this material.
As it is well known \cite{bs}, after twisting, the
Lagrangian of $N=2$ SYM is invariant under
\beqa
\label{BRST} 
&&sA=\psi-Dc\ \
,\nonumber\\ &&s\psi=-[c,\psi]-D\phi\ \ ,\nonumber\\
&&s\phi=-[c,\phi]\ \ ,\nonumber\\ 
&&sc=-{1\over 2}[c,c]+\phi.
\eeqa
The distinction between the cases with a vacuum expectation value
of the scalar field equal or different to zero, which was important
for the discussion in \cite{bftt}, is of no relevance here. We
will not dwell on this subject anymore, working with (\ref{BRST})
which is the simplest set of equations. The BRST operator, $s$, defined 
in (\ref{BRST}) is
such that $s^2=0$ and when
the set of equations in (\ref{BRST}) is restricted
to the solutions of the
Euler--Lagrange classical equation ({\it the zero modes})
it gives the derivative on the space $\mathscr{M}^{+}$.
In (\ref{trecinque}) and (\ref{f.9}) we already described these 
solutions
in terms of the parameters of the ADHM construction for $A$ and
$\psi$. The form of the other fields appearing in (\ref{BRST})
is \cite{bftt}
\beqa
c&=&U^\dagger s U, \\
\phi&=& U^{\dagger}{\cal M} f {\cal M}^{\dagger}U+U^{\dagger}{\cal A} U.
\label{restoeq}
\eeqa
Plugging (\ref{trecinque}) and (\ref{f.9}) into (\ref{BRST}) leads
to the action of the operator $s$ on the elements of the
ADHM construction
\beqa
{\cal M}&=&s\Delta+C\Delta={\cal S}\Delta\label{azzarolina1}\ \ ,\\
{\cal A}&=&s{\cal M}\Delta+C{\cal M}={\cal S}{\cal M}\label{azzarolina2}\ \ ,\\
s{\cal A}&=&-[C,{\cal A}]\label{azzarolina3}\ \ ,\\
sC&=&{\cal A}-C C\label{azzarolina4}\ \ ,
\eeqa
{\it i.e.} this is the realization of the BRST algebra on 
the instanton moduli space. $C$ is the connection we have introduced
in (\ref{tre}). In the $k=2$ case, using (\ref{a}), (\ref{azzarolina1})
can be written as
\beq
({\widetilde{\cal M}}_{\alpha\dot{\alpha}})_i=
\sigma^{\mu}_{\alpha\dot{\alpha}}(K_{\mu\nu})_{ij}
(s\widetilde{\Delta}_{\nu})_j.
\label{areale}
\eeq
We can now use (\ref{azzarolina1}) to further elaborate on 
the results of the previous section where we constructed
hyperk\"ahler metrics. We
begin by reminding that the K\"ahler potential 
for those varieties is given by the second moment of the gauge
field strength distribution \cite{maciocia}
\beq
{\cal K}=\int_{\bR^4}x^2 |F|^2={1\over 2}\Tr \left[a^\dagger
(1+P_\infty)a \right],
\label{calpot}
\eeq
where $P_\infty=\lim_{x\to\infty}P=1 - b b^\dagger$ 
and $P=U(x)U(x)^\dagger$ is the projector onto $Ker\Delta^\dagger$.
Defining complex derivatives by $s=\partial+\bar\partial$ the
K\"ahler form is given by
\beq
\omega_{\mathscr{M}^{+}}=\partial\bar\partial{\cal K}=
{1\over 2}\Tr \left[{(\cal S}a)^\dagger
(1+P_\infty){\cal S}a \right],
\label{kalformbos}
\eeq
with ${\cal S}=s+C$ the covariant derivative coming 
from (\ref{azzarolina1}). In the $k=2$ case, substituting (4.14) 
of \cite{bftt} in (\ref{kalformbos}), 
one recovers (\ref{metric1}).

Before closing this section we remind the reader that
in the topological formalism of ref.\cite{bftt} the measure arises
as a consequence of (\ref{azzarolina1}). Let us see how.
At the semi--classical level, any correlator which is expressed
as a polynomial in the fields, becomes after projection
onto the zero--mode subspace, a well--defined  
differential form on $\mathscr{M}^+$ \cite{witten}.
Symbolically
\beq
\label{prescription}
\left< fields \right> = \int_{\mathscr{M}^{+}}  \ 
\left[ (fields)\ e^{-S_{\rm TYM}} \right]_{zero-mode\  subspace}
\ \ .
\eeq

Let us now call $\{ \widehat{\Delta}_{i}\}$ ($\{\widehat{\cal M}_i\}$), 
$i=1,\ldots, p$, where $p=8k$, a basis of (ADHM) coordinates on 
$\mathscr{M}^+$ ($T_A\mathscr{M}^+$).
(\ref{azzarolina1}) thus yields 
$\widehat{\cal M}_i = s \widehat{\Delta}_i + (\widehat{{\cal C} \Delta})_i$. 
A generic  function on the zero--mode subspace
will then have the expansion
\beqa
\label{inte1}
g( \widehat{\Delta}, \widehat{\cal M}) & = & 
g_{0} (\widehat{\Delta} )+ 
g_{i_1} (\widehat{\Delta} ) \widehat{\cal M}_{i_1} + 
{1\over 2!}
g_{i_1 i_2}(\widehat{\Delta}) \widehat{\cal M}_{i_1}\widehat{\cal M}_{i_2} +
\ldots 
\nonumber \\
&+&
{1\over p!}
g_{i_1 i_2 \ldots i_p} (\widehat{\Delta}) 
\widehat{\cal M}_{i_1}\widehat{\cal M}_{i_2}\cdots \widehat{\cal M}_{i_p}
\ \ ,
\eeqa
the coefficients of the expansion being totally antisymmetric in their indices.
Now (\ref{azzarolina1}) 
implies that the
$\widehat{\cal M}_{i}$'s and the $s\widehat{\Delta}_{i}$'s are related 
by a (moduli--dependent) linear transformation $K_{ij}$, 
which is completely known 
once the explicit expression for ${\cal C}$ is 
plugged into the $\widehat{\cal M}_{i}$'s:
\beq
\label{inte2}
\widehat{\cal M}_{i} = K_{ij} ( \widehat{\Delta} ) s\widehat{\Delta}_{j}
\ \ .
\eeq
It then follows that 
\beqa
\label{noncera}
\widehat{\cal M}_{i_1}\widehat{\cal M}_{i_2}\cdots \widehat{\cal M}_{i_p} &=&
K_{i_1 j_1} K_{i_2 j_2} \cdots K_{i_p j_p} 
s\widehat{\Delta}_{j_1}s\widehat{\Delta}_{j_2}\cdots s\widehat{\Delta}_{j_p} =
\nonumber \\
&=& \epsilon_{j_1 \ldots j_p} K_{i_1 j_1} K_{i_2 j_2} \cdots K_{i_p j_p} 
\ s^p \widehat{\Delta} =
\nonumber \\
&=&
\epsilon_{i_1 \ldots i_p} ({\rm det} K) \  s^p \widehat{\Delta}
\ \ ,
\eeqa
where $s^p \widehat{\Delta}\equiv 
s\widehat{\Delta}_1 \cdots s \widehat{\Delta}_p$.
From (\ref{inte1}), (\ref{inte2}) we conclude that
\beqa
\int_{\mathscr{M}^+}g( \widehat{\Delta}, \widehat{\cal M}) &=& 
{1\over p!} \int_{\mathscr{M}^+} g_{i_1 i_2 \ldots i_p} (\widehat{\Delta} ) 
\widehat{\cal M}_{i_1}\widehat{\cal M}_{i_2}\cdots \widehat{\cal M}_{i_p} =
\nonumber \\
&=&
\int_{\mathscr{M}^+}  s^p  \widehat{\Delta} \  |{\rm det} K| 
g_{1 2 \ldots p} (\widehat{\Delta} )
\ \ .
\label{finprescription}
\eeqa
The determinant of $K$ naturally stands out as 
{\it the instanton integration measure for $N=2$ SYM theories.} 
This important ingredient of the calculation
is obtained in standard instanton calculations as
a ratio of bosonic and fermionic zero--mode Jacobians,
while it emerges here in a geometrical and  very direct way.

\section{A Supermanifold Construction\label{quattro}}
\setcounter{equation}{0}
The purpose of this section is to show that the results of the
previous sections can be consistently put together into
a supermanifold framework. 
We do this into two steps. In the first
subsection we recall some general notions on supermanifolds, which we use
in the second subsection to do the construction itself.
\subsection{Generalities on Supermanifolds\label{subgen}}
In this subsection we give the definition of a supermanifold and
show that these objects can  conveniently be constructed from vector
bundles. Naively, a supermanifold is a manifold with both ``commuting
and anticommuting'' coordinates. One possible mathematical
formalization of this idea is provided by the so-called
Berezin-Le\u\i tes-Kostant approach \cite{BL,K}, where one considers
an ordinary differentiable manifold $X$ and ``enlarges'' its
structure sheaf (the sheaf of germs of $C^\infty$ functions on 
$X$)\footnote{For the definition of the notion of sheaf
see e.g. \cite{G,BBH}.}
to a sheaf $\A$ including anticommuting generators.

Let us recall a few algebraic facts.
A $\Z_2$-graded commutative
algebra $\Lambda$ is an associative unital algebra $\Lambda$
over the real field $\real$ which has a splitting
$\Lambda=\Lambda_0\oplus\Lambda_1$ such that
\beq ab=(-1)^{\alpha\beta}ba\quad\mbox{if}\quad
a\in \Lambda_\alpha,\ b\in\Lambda_\beta.\eeq
The field $\real$ is embedded into $\Lambda$ by $x\mapsto x\cdot 1$.
A morphism $\phi\colon\Lambda\to\Lambda'$
between two such algebras is an algebra morphism which
is compatible with the grading, i.e., $\phi(\Lambda_\alpha)\subseteq
\Lambda'_\alpha$.

An $(m,n)$-dimensional
supermanifold is a pair $\X=(X,\A)$, where $X$ is an $m$-dimensional
differentiable manifold, and $\A$ is a sheaf of $\Z_2$-graded commutative
algebras on $X$,
satisfying the following requirements:

\begin{itemize}\item
if $\cal N$ is the nilpotent\footnote{An element in an algebra
is nilpotent if it vanishes when raised to a finite power.}
subsheaf of $\A$, then
$\A/\cal N$ is isomorphic to the sheaf ${\cal C}_X^\infty$
of $C^\infty$ functions on $X$. The quotient map
$\sigma\colon\A\to{\cal C}_X^\infty$ is often called the \emph{body map}.
\item Locally the sheaf $\A$ is a sheaf of exterior algebras
over the smooth functions with $n$ generators; namely, every $x\in X$ has
a neighbourhood $W$ such that there is an isomorphism
$\A(W)\simeq C^\infty_X(W)\otimes\wedge V$, where $V$
is an $n$-dimensional vector space. This isomorphism
is required to be compatible with the map $\sigma$.
\end{itemize}
If $(x^1,\dots,x^m)$ are local coordinates in $W$,
and $(\theta^1,\dots,\theta^n)$ is a basis of $V$, the collection
$(x^1,\dots,x^m,\theta^1,\dots,\theta^n)$
is said to be a local coordinate chart for $\X$.
According to the second requirement above,
a local section of $\A$ (i.e., a superfunction on $\X$) has a local expression
\begin{eqnarray}
f &= & f_0(x)+\sum_{\alpha=1}^nf_\alpha(x)\,\theta^\alpha \nonumber\\
&+&\sum_{\begin{array}{c} \\[-26pt]\mbox{\tiny $\alpha,\beta=1,\dots
n$} \\[-10pt] \mbox{\tiny $\alpha<\beta$}\end{array}}
f_{\alpha\beta}(x)\,\theta^\alpha\theta^\beta+\dots+f_{1\dots n}(x)
\theta^1\cdots\theta^n.\label{superfield}\end{eqnarray}
This is quite evidently the physicists' superfield expansion.
The map $f\mapsto f_0$ is the coordinate expression of the
map $\sigma\colon\A\to{\mathcal C}_X^\infty$.

Supervector fields and differential superforms may be introduced in terms
of the notion of graded derivation of a graded commutative algebra $\Lambda$.
A homogeneous graded derivation $D:\Lambda\to\real$ is a linear map
satisfying a graded Leibniz rule
\beq
D(ab)=D(a)b+(-)^{\alpha|D|}aD(b),
\eeq
where $a\in\Lambda_\alpha$ and $|D|=0, 1$. $D$ is said to be even (odd)
if $|D|=0$ ($|D|=1$). A graded derivation is the
sum of an even and an odd homogeneous graded derivation. The space of such
graded derivations will be denoted by $\mbox{Der}_\bR\Lambda$.
The sheaf of derivations, $\mbox{Der}\A$, of the sheaf of superfunctions
is defined by the rule $(\mbox{Der}\A)(W)=\mbox{Der}_\bR\A(W)$ for any open
set $W\subseteq X$. This is the sheaf of sections of supervector bundle on
$\X$ of rank $(m,n)$, called the tangent superbundle to $\X$; its sections are
the supervector fields. The sections of the dual superbundle are the
differential 1-superforms and by taking graded wedge products one defines the
sheaves $\Omega^k_{\X}$ of differential $k$-superforms. In the local
coordinate charts introduced above a $k$-superform is written as
\beq
\omega=\sum_{\begin{array}{c} \\[-26pt]\mbox{\tiny $p+q=k$}
\\[-10pt] \mbox{\tiny $i_1,\dots ,i_p=1,\dots ,m$}
\\[-10pt] \mbox{\tiny $\beta_1,\dots ,\beta_q=1,\dots ,n$}
\end{array}}  \omega_{i_1\dots i_p,\beta_1\dots\beta_q}(x,\theta)\ 
dx^{i_1}\wedge\dots\wedge dx^{i_p}\wedge d\theta^{\beta_1}\wedge\dots\wedge
d\theta^{\beta_q}.
\eeq

For later use, we recall that a {\it supermetric} on $\X$ is a graded-symmetric
nondegenerate pairing $T\X\otimes T\X \to \A$. In local
coordinates $(x,\theta)$ a supermetric is written as
\beq
g=g_{ij}\,dx^i\otimes
dx^j+g_{i\alpha}dx^i\otimes
d\theta^\alpha+g_{\alpha i}d\theta^\alpha\otimes
dx^i+g_{\alpha\beta}\,d\theta^\alpha\otimes d\theta^\beta\label{genmetr}\eeq
where the matrix of superfunctions $g_{ij}$ 
is symmetric and $g_{\alpha\beta}$ is skew-symmetric.

We study now the relation between supermanifolds and vector bundles.
Given a rank $n$ vector bundle $E$ on $X$,
the pair $(X,\A)$, where $\A$ is the sheaf of sections ${\mathcal C}^\infty
(\wedge E)$
of the exterior algebra bundle\footnote{The exterior bundle
$\wedge E=\oplus_{i=1}^n\wedge^i E$ is the direct sum of the antisymmetrized
tensor product of $i$ copies of $E$.} $\wedge E$, is a supermanifold.
Indeed, if ($\theta^1,\dots,\theta^n$) is  a local basis of sections of $E$,
a section  of ${\mathcal C}^\infty(\wedge E)$ has the form of (\ref{superfield})
(so the sections of $E$ are regarded as Grassmann variables and the sections of the
dual bundle, $E^*$, are fermion fields).
As a matter of fact all supermanifolds
are of this kind; to extract the data corresponding to a vector
bundle from a supermanifold, heuristically we may regard the odd
coordinates $\theta^\alpha$ as the basis sections which locally generate the bundle.
Intrinsically, with no reference to a coordinate system,
if $(X,\A)$ is an ``abstract'' supermanifold,
and $\mathcal N$ is the nilpotent subsheaf of $\A$, then
the quotient sheaf ${\mathcal N}/{\mathcal N}^2$
is the sheaf of sections of a vector bundle $E$, and 
$\A\simeq {\mathcal C}^\infty
(\wedge E)$. This fact is known as \emph{Batchelor's theorem}
\cite{Batch}.

If a supermanifold $\X=(X,\A)$ is represented by a vector bundle $E$,
its tangent superbundle $T\X=T_0\X\oplus T_1\X$ is explicitly described by the
isomorphisms
\beq
T_0\X\simeq\A\otimes TX,\qquad T_1\X\simeq \A\otimes \E^\ast
\label{supertg}
\eeq
where $\E$ is the sheaf of sections of $E$.
Indeed, the derivations $\partial/\partial x^i$ locally generate $TX$,
while, in view of the relation
\beq
\frac\partial{\partial \theta^\beta}\theta^\alpha=\delta^\alpha_\beta,
\label{dualita}
\eeq
the odd derivations $\partial/\partial\theta^\alpha$ may be regarded
as local generators for $\E^*$ (the sheaf of sections of the dual bundle to $E$).
This also shows that there is a map $T_1^*\X\hookrightarrow\A$ given by
$d\theta^\alpha\mapsto\theta^\alpha$. 

With this explicit representation of a supermanifold, a natural way
to introduce a (block-diagonal) supermetric $\gamma$ on it is to assign a
Riemannian metric $g$ for the even sector, and a  nondegenerate alternate
two-form $\chi$ on $E$ (notice that the matrix of the bosonic metric must 
be symmetric while the matrix of the fermionic metric has to be 
skew-symmetric); with reference to (\ref{supertg}), we  have
\beq
\gamma(u_0+u_1,v_0+v_1)=g(u_0,v_0)+\chi(u_1,v_1).
\eeq

As a particular case $E$ may be the tangent or cotangent bundle to $X$,
in which case $m=n$. If $E=T^*X$, an isomorphism
$\X\simeq (X,{\mathcal C}^\infty(\wedge T^\ast X))$
may be locally expressed in the form
$\theta^i=\eta^i$, where the $\theta$'s are odd coordinates
on $\X$, and $(\eta^1,\dots,\eta^m)$ form a basis of sections of the
cotangent bundle.

Suppose now that $X$ is a hermitian manifold with hermitian metric $g$;
considering also the associated two-form $\omega$, we have the data to
define a block-diagonal supermetric:
\beq
\gamma=g_{ij}\,dx^i\otimes dx^j+\omega_{\alpha\beta}\,d\theta^\alpha\otimes
d\theta^\beta.
\eeq
So, let us
consider the supermanifold $\X=(X,\A)$, where $\A={\mathcal
C}^\infty(\wedge T^\ast X)$ (here $T^\ast X$ is the
complexified $C^\infty$ cotangent bundle). Let $(\eta^1,\dots,\eta^n)$ be
linearly independent forms of type $(1,0)$. Then locally we have
\beq g=g_{i\bar\jmath}\,\eta^i\otimes\bar\eta^{\bar\jmath}\eeq
and the matrix $g_{i\bar\jmath}$ is real and skew-symmetric. 

The 1-forms $(\eta^1,\dots,\eta^n,\bar\eta^1,\dots,\bar\eta^n)$
provide local odd coordinates for $\X$, and with all these data,
and with the help of the dual of (\ref{supertg}),
we may define a supermetric $\gamma$ for $\X$, by letting
\beq
\gamma=g_{i\bar\jmath}\,\eta^i\otimes\bar\eta^{\bar\jmath} +i
g_{i\bar\jmath}\,d\eta^i\otimes d\bar\eta^{\bar\jmath}.
\label{supermetric}
\eeq
So we use the datum provided by the specification of the
1-forms $\eta^i$ to ``replicate'' the (even) metric $g$
in the odd sector of the supermanifold. Notice that the bosonic metric
is hermitean while the fermionic one is skew-hermitean.
We find now convenient to use real coordinates to compute the 
superdeterminant of $\gamma$, by introducing $\theta^i=\Re\eta^i,
\theta^{i+n}=\Im\eta^i$. To simplify the notation we will use the same
symbol $g_{ij}$ also to denote the metric in real coordinates.
Suppose now that $(x^1,\dots,x^{2n})$ are real coordinates for $X$.
Let $K$ be the matrix of the components of the 1-forms
\beq
\theta^i=K_{ik}dx^k,
\label{coordchg}\eeq
over the basis $(dx^k)$.
A simple computation shows that in the supercoordinates
$(x^1,\dots,x^{2n},$\break
$\theta^1,\dots,\theta^{2n})$
one has
\beq \mbox{Sdet}\,\gamma = (\mbox{det}\, K)^2.\label{sdet}\eeq

Assume now further that $X$ is hyperk\"ahler, i.e.,
$X$ has a hermitian metric $g$ with three compatible
complex structures $J_i$ which generate
the quaternion algebra (that is, $\nabla J_i=0$
where $\nabla$ is the Levi-Civita connection of $g$,
and $J_i\,J_h=-\delta_{ih}+\varepsilon_{ihk}\,J_k$).
The same definition --- mutatis mutandis --- 
applies to a complex supermanifold $(X,\A)$
endowed with a hermitian supermetric.
Let us still consider the case where $\A$ is the sheaf of sections
of the exterior algebra of $T^\ast X$, and 
fix 1-forms $\eta^j$ as before. Since 
equations (\ref{supertg}) now reads
\beq T\X\simeq\A\otimes (TX\oplus TX),\eeq
the three basic complex structures of $X$ can be
lifted to $\X$ (the matrices of the complex structures
in the odd sector expressed on the basis $(d\eta^i,d\bar\eta^{\bar\jmath})$
are the same as the matrices of the complex structures in the even sector
expressed on the basis $(\eta^i,\bar\eta^{\bar\jmath})$), and these are 
automatically compatible with the supermetric of $\X$. Thus,
\emph{the supermanifold $\X$ acquires a hyperk\"ahler structure.}

\subsection{Supermanifolds by (even) group quotients}
We reconsider now the construction in Section \ref{quotients}
in the case of supermanifolds. The basic theory of the
action of a super Lie group $\mathfrak G=(G,{\cal H})$ on a supermanifold 
$\X=(X,\A)$ was developed
in \cite{K}. This theory is quite involved because
the geometry of a supermanifold is not completely encoded
in the underlying topological manifold but, of course, it
is also described by the structure sheaf (sheaf of superfunctions).
As a result one needs to formulate the theory in purely sheaf-theoretic
terms (see e.g.~\cite{BBH}) or using the graded Hopf algebras
of global functions on the supergroup \cite{K}. Here we shall
only notice that, provided that some conditions for the existence
of a good quotient are satisfied, the action of $\mathfrak G$ on
$\X$ defines a quotient supermanifold $\mathfrak Y=(Y,\B)$ such that
$\dim\mathfrak Y=(m-p,n-q)$ if $\dim \X=(m,n)$ and $\dim \mathfrak G=(p,q)$.
As a consequence of a result proved in \cite{austr}, these conditions
are equivalent to the fact that the induced action of the bosonic
part of the group $G$ on $X$ yields a good quotient $Y$.

The case of interest to us is simpler than the general 
situation for two reasons. The first is that
 the symmetry group $G$ is purely bosonic,
so that the dimension of the quotient is $(m-r,n)$ if $\dim G=r$.
This simplification is present for any number $N$ of
supersymmetries. The second fact is typical of $N=2$;
in this case the dimension of the bosonic moduli space
equals the number of fermionic zero-modes, which
may be interpreted as differential 1-forms on the moduli space.
From the viewpoint of the mathematical description we propose here
this means that the vector bundle associated with the
structure sheaf of the supermoduli space is the cotangent
bundle. When this happens, one can get an action of $G$ on 
$\X$ from an action of $G$ on the bosonic manifold
$X$; indeed, the latter induces by linearization an action on $T^\ast X$
(i.e., on the $\theta$'s)
which is then extended to an action on $\A$.

The Marsden-Weinstein reduction procedure now works
as follows. Let $\V=(V,\A)$ be an $(s,q)$-dimensional supermanifold with a
supermetric $g$. Assume also that in some local coordinate system
$(x^1,\dots, x^s,\theta^1,\dots,\theta^q)$ the supermetric
has a block diagonal form,
\beq \gamma=g_{ij}\,dx^i\otimes dx^j+g_{\alpha\beta}\,d\theta^\alpha\otimes
d\theta^\beta.\eeq
If $G$ has dimension $r$, the quotient supermanifold
$\M=(M,\B)$ has dimension $(m,q)$ with $m=s-r$. 
The ``body'' manifold $M$ is the quotient $V/G$.
The connection $C$ is ``purely even,'' in the sense that, locally,
$C=C_i\,dy^i$ where $(y^1,\dots,y^m)$ are even local coordinates on $M$.
The quotient metric has the form
\beq\tilde \gamma = \tilde g_{ij} \,dy^i\otimes
dy^j+g_{\alpha\beta}\,d\theta^\alpha\otimes d\theta^\beta
\label{supquotmet}\eeq
where the even components $\tilde g_{ij}$ are given by equation
(\ref{metric}). 

The square root of the superdeterminant of $\tilde\gamma$ 
gives the supermeasure of the supermoduli space for any
number $N$ of supersymmetries, since the latter specifies
the ``odd'' geometry of the supermoduli space. 
For $N=2$ this construction can be further specialized. The supermetric, 
for instance, can be computed using (\ref{supermetric}), and the
results of sections \ref{quotients} and \ref{topological} can be given
a simple description in terms of supermanifold theory.
To do so, we need to introduce a dictionary between the general supermanifold
theory and the ADHM construction of the moduli space of instantons.
We consider a supermanifold $\M=(\mathscr{M}^+,\B)$
whose bosonic part $\mathscr{M}^+$ is the moduli space of
section \ref{topological}. 
The fermionic zero-modes, according to
(\ref{azzarolina1}), can be interpreted as differential 1-forms
on $\mathscr{M}^+$. It is therefore natural to construct
the supermanifold, $\M$,  as explained previously
in this section, in terms of the cotangent bundle $T^\ast\mathscr{M}^+$,
i.e, $\B={\cal C}^\infty(\wedge T^\ast\mathscr{M}^+)$.
Thus, $\dim\M=(8k,8k)$. 
Since $M$ is hyperk\"ahler, we get a hyperk\"ahler
structure for the supermoduli space $\M$. 
In the $k=2$ case (\ref{supermetric}) is realized using the
coordinates employed  in (\ref{piattatg1}) while (\ref{coordchg}) 
is given by (\ref{areale}). The supermeasure is straightforwardly
obtained by (\ref{sdet}) which coincides with the one found  in \cite{bftt}.
This shows how the supermeasure can be obtained by a superquotient
construction. In the next section we will show how to implement
this construction in the functional integral which computes
the correlators of interest for generic values of $k$. 
Following \cite{maciocia}
we can also write a potential for the hyperk\"ahler supermetric:
\beq f=a^\dagger(1+P_\infty)a+\bar\mathcal M(1+P_\infty)\mathcal M.
\label{superpot}\eeq
Indeed, by acting on (\ref{superpot}) with the operator 
$\partial\bar\partial$, where $\partial$ is the holomorphic 
exterior differential on $\M$, it is easy to recover the K\"ahler
superform associated with the supermetric $\tilde\gamma$.

\section{ The Hyperk\"ahler Quotient Construction and Multi-Instanton 
Calculus\label{secfin}}
\setcounter{equation}{0}
In section \ref{topological} we have obtained the explicit form 
(\ref{finprescription}) of the instanton
dominated correlators by plugging the constraints
(\ref{natale}), (\ref{snatale}), (\ref{f.16}), (\ref{azzarolina1})
into (\ref{prescription}). This is possible only for winding numbers
up to $k=2$ for which an explicit solution to the contraints is known.
For arbitrary winding numbers an explicit solution to (\ref{a.2}) and
(\ref{fconstr}) (and, as a consequence, to (\ref{azzarolina1})) is
missing. It can be 
useful though, as it was shown in \cite{DKM2} for the $N=4$ case, to have
an expression of the type (\ref{prescription}) in which the manifold 
on which to perform the integration is given by the unconstrained ADHM
parameters introduced in (\ref{salute}) and the constraints
are introduced by suitable Dirac deltas. In the bosonic case, which we
treat first as an example, this program goes into the opposite direction of 
the strategy that we have adopted starting from (\ref{piatta}) to end with
(\ref{metric1}). With respect to the computations carried on in \cite{DKM1},
we do not have to worry about the transformation properties of our measure
which is built to be hyperk\"ahler (see (\ref{superpot})) and 
thus $s$-invariant.
 
At first we will stick to the $k=2$ case since, as we shall see later,
the extension to arbitrary winding numbers is straightforward. 
In order to make a bridge between the ADHM construction and the computations
performed in section \ref{quotients} let us remind the reader that the 
connection $C$ introduced in (\ref{azzarolina1}) is a matrix of the form
\beq
C =\pmatrix{0 & 0\cr 0 &  C^{12}\cr  - C^{12} & 0 }
\ \ 
\eeq
Plugging (\ref{azzarolina1}) into (\ref{fconstr}) leads to 
\beq
\Delta^{\dagger} C\Delta-\bigl(\Delta^{\dagger} C\Delta\bigr)^T
=(\Delta^{\dagger}s\Delta)^T-\Delta^{\dagger}s\Delta\ \ ,
\label{C-eq}
\eeq
or, more compactly, to 
\beq
L\cdot C = -  \Lambda_C \ \ , 
\label{cipreq}
\eeq
and to an  explicit 
expression for $C^{12}$ which is equal to  (\ref{pizza1}) \cite{bftt}.
Let us now make the sigma-model interpretation of the computations in
section \ref{quotients} more explicit. The starting point is the
metric (\ref{metric1}) which can be interpreted \cite{hklr}
as the target space metric (described by the coordinates 
(\ref{coordv})) arising from the Lagrangian
\beqa
L&=&g^{\mathscr{M}^{+}}_{AB}sm^A sm^B=\bigg(g^{\mathscr{N}^{+}}_{AB} - 
{{g_{AC}^{\mathscr{N}^{+}}g_{BD}^{\mathscr{N}^{+}}k^Ck^D}
\over {g_{EF}^{\mathscr{N}^{+}}k^Ek^F}}\bigg)sm^A sm^B\nonumber\\&=&
g^{\mathscr{N}^{+}}_{AB}sm^A sm^B-C^{12}C_{12}=g^{\mathscr{N}^{+}}_{AB}
{\cal S}m^A {\cal S}m^B
\label{lagran}
\eeqa
where ${\cal S}m^A=(sm^A+C^{12}k^A)$ is the covariant derivative on 
$\mathscr{M}^{+}$.
These formulae are derived from (\ref{tre}) and (\ref{metric}) in the case
where the Lie algebra $\g=SO(2)$. In this particular case the metric
$g^{ab}$ is a one by one metric and $C^{12}=H C_{12}$\footnote{The 
connection $C^{12}$ appearing in (\ref{lagran}) is a one-form on
the space spanned by 
the coordinates (\ref{coordv}). The coordinate $a_1$ 
in (\ref{pizza1}) is to be replaced by its value (\ref{natale}).}.

Imposing the constraint (\ref{natale}), the change of variable (\ref{snatale})
and the explicit form  (\ref{aripizza}) is now straightforward. Following 
\cite{DKM1} we write
\beq
1=16\vert a_3\vert^2\int \delta ({1\over 4} tr_2 \sigma^a 
(\Delta^\dagger\Delta-(\Delta^{\dagger}\Delta)^{T})\delta(\bar a_3 a_1\bar 
a_1 a_3-{\Sigma\over 2}),
\label{delta1}
\eeq
For consistency with the other sections where we constantly used forms,
to impose the constraints we can use currents \cite{dr} instead of Dirac 
deltas. In fact the definition of an n-current  
\beq
\int T=\int \sum_{i_1<\ldots<i_n}T_{i_1\ldots i_n}dx_1\wedge\ldots\wedge dx_n,
\eeq
encompasses also the case where $T_{i_1\ldots i_n}$ is a 
distribution. So from now on when we write expressions like (\ref{delta1}) we
shall omit the differentials. 
Taking these observations into account, we can 
write
\beqa
\int e^{-L}&=&\int \int dC^{12}\delta(C^{12}-{1\over H}
\Bigl( w_1^\mu dw_2^\mu -
w_2^\mu dw_1^\mu - 4a_1^\mu da_3^\mu + 
{{d\Sigma}\over 2}\Bigr)\nonumber\\
&&16\vert a_3\vert^4\delta ({1\over 4} tr_2 \sigma^a 
(\Delta^\dagger\Delta-(\Delta^{\dagger}\Delta)^{T})\delta(\bar a_3 a_1\bar 
a_1 a_3-{\Sigma\over 2})e^{-L^\prime},
\label{fibos}
\eeqa
where the integration is over the set of variables (\ref{coordnv}),
\beq
L^\prime={\cal S}m^I{\cal S}m^I=
({\cal S}a)^\dagger(1+P_\infty){\cal S}a=sm^Ism^I-C C H,
\label{bosnv}
\eeq
and $C$ is given by (\ref{a}). This shows that the K\"ahler form (\ref{bosnv})
descends on the quotient manifold (\ref{lagran}).

The extension to the case of (\ref{prescription}) is now straightforward.
When projected onto the zero-modes subspace of winding number $k$, 
$S_{\rm TYM}=\left[ S_{\rm inst}\right]_{k}= 
\left[S_B\right]_{k} + \left[S_F\right]_{k}$ and
\beqa
\label{abm}
\left[ S_{B}\right]_{k}&=&
4\pi^2\Tr\bigg[
2(\bar v v) \sum_{l=1}^{k}|w_{l}|^2+
\sum_{l,p=1}^{k}(\bar{w}_{l}\bar v
w_{p}-\bar{w}_{p}\bar v w_{l})({\cal A}^{\prime}_b)_{lp}\bigg]
\ \ ,
\\
\label{afm}
\left[ S_{F}\right]_{k}&=&
4\pi^2\Tr\bigg[ -2\bar v
\sum_{l=1}^{k}\mu_{l}\bar{\mu}_{l}+
\sum_{l,p=1}^{k}(\bar{w}_{l}\bar v
w_{p}-\bar{w}_{p}\bar vw_{l})({\cal A}_{f}^{\prime})_{lp}\bigg]
\ \ ,
\eeqa
${\cal A}_{b}^{\prime}$ and ${\cal A}_{f}^{\prime}$ being defined as
\beq
\label{decca}
{\cal A}^{\prime}= {\cal A}^{\prime}_b + {\cal A}^{\prime}_f 
\ \ ,
\eeq
where 
\beqa
\label{decca1}
L\cdot{\cal A}^{\prime}_b 
&=& - \Lambda_{b} (v) \ \ ,
\\ 
\label{decca2}
L\cdot{\cal A}^{\prime}_f &=& - \Lambda_f.
\eeqa 
The right hand sides of (\ref{decca1}) and (\ref{decca2})
are given by \cite{bftt}
\beqa
\label{lb-def}
[\Lambda_{b}]_{ij} (\Omega_0)& =&  
\bar{w}_i \Omega_0 w_j - \bar{w}_j \Omega_0 w_i,\\
\Lambda_f &=& {\cal M}^{\dagger}{\cal M} - ({\cal M}^{\dagger}{\cal M})^T 
\eeqa
and
\beq
\lim_{|x| \rightarrow \infty}\phi \equiv 
\lim_{|x| \rightarrow \infty} U^\dagger{\cal A} U =
v \frac{\sigma_{3}}{2i} \ \ .
\eeq
Now $\left[S_B\right]_{k}$ contains only the $w_l$ variables which are 
unconstrained, while
\beq
\label{yukact}
[S_F]_k=
\overline{{\cal M}}_{i}^{\dot{A}\alpha}
(h_{ij})_{\alpha}{}^{\beta}
({\cal M}_{j})_{\beta\dot{A}}=({{\cal S}}_{i}^{\dot{A}\alpha}a)^\dagger
(h_{ij})_{\alpha}{}^{\beta}
({\cal S}_{j})_{\beta\dot{A}}a,
\eeq
where $i,j=1,\ldots,n$ and%
\footnote{In the following equation  we denote by $h^\dagger$ 
the hermitean conjugate matrix  obtained {\it without} complex conjugating
$v$, {\it i.e.} treating $v$ as real.}
$h=-h^\dagger$.
(\ref{yukact}) is of the same form of (\ref{lagran}), and leads to  
(\ref{finprescription}) with $g_{1 2 \ldots p}=det (h)$. The difference
from the bosonic case (\ref{lagran}) is given by the fact that 
after expanding the fermionic action, the measure
arises from the change of variables from ${\cal S}\Delta$ to $s\Delta$,
as a consequence of (\ref{sdet}).

To put (\ref{yukact}) in a form similar to (\ref{bosnv}),
in addition to the constraints introduced in (\ref{fibos})
we also have to insert the following deltas to take care of the fermionic 
constraint (\ref{fconstr}), the BRS relation (\ref{azzarolina1}) and 
the presence of a scalar field in the action
\beqa
1&=&{1\over 16\vert a_3\vert^4}\int d^4M_1\delta((\Delta^\dagger{\cal M})-
(\Delta^\dagger{\cal M})^T),\nonumber \\
1&=&\int d\mu_1\delta(\mu_1-{\cal S}w_1),\nonumber\\
1&=&\int d\mu_2\delta(\mu_2-{\cal S}w_2),\nonumber\\
1&=&\int dM_3\delta(\mu_1-{\cal S}a_3),\nonumber\\
1&=&H\int d{\cal A}_{12}\delta(\Delta^\dagger{\cal A}\Delta-
(\Delta^\dagger{\cal A}\Delta)^T).
\label{delta2}
\eeqa
Consequently (\ref{prescription}) gets modified to
\beqa
\label{prescription1}
&&\left< fields \right> = \int_{\mathscr{M}^{+}}  \
dC d{\cal A}\delta(C+L^{-1}\Lambda_C )
\delta(\Delta^\dagger{\cal A}\Delta-
(\Delta^\dagger{\cal A}\Delta)^T)
\delta((\Delta^\dagger{\cal M})-
(\Delta^\dagger{\cal M})^T)\nonumber\\
&&
\delta({\cal M}-{\cal S}\Delta)\delta ({1\over 4} tr_2 \sigma^a 
(\Delta^\dagger\Delta-(\Delta^{\dagger}\Delta)^{T})\delta(f(C))
\left[ (fields)\ e^{-S_{\rm TYM}} \right]_{zero-mode},
\ \ 
\eeqa
In the $k=2$ case, $f(C)=C\cdot n=\bar a_3 a_1\bar+ a_1 a_3-\Sigma/2$
where $n^A=(0,0,a_3/2,\Sigma)$ is a certain direction in the moduli space.

Rotations of the type (\ref{rmatrix}) act on $f(C)$ so that
\beq
1=\triangle_f(C)\int d\theta \delta(f(C^\theta)),
\label{fp}\eeq
where 
\beq
\triangle_f(C)={\delta f(C)\over\delta\theta}=\left|
|a_3|^2-|a_1|^2-\left.\frac{1}{8}\frac{\partial\Sigma^{\theta}}
{\partial\theta}
\right|_{_{\theta=0}}\right|,
\eeq
as found in \cite{DKM1}. 
$\triangle_f(C)$ is invariant under $O(2)$ rotations. 
The reader will recognize in (\ref{fp}) the
standard Faddeev-Popov trick. Then after multiplying (\ref{prescription1})
by ${2\pi}^{-1}\int d\theta$ and expanding the forms in the usual coordinate
basis, $\delta(f(C))$ can be made to disappear using (\ref{fp}).

\section*{Acknowledgements}
This work was supported in part by the EEC contract HPRN-2000-00122, by
the INTAS project 991590 and by Friuli-Venezia Giulia Region
through the research project ``Noncommutative geometry: algebraic,
analytical and probabilistic aspects and applications to mathematical 
physics.'' One of the authors (F.F.)
wants to thank D.Bellisai for collaboration in an early stage
of this work and G.C.Rossi for carefully reading the manuscript.

\end{document}